\documentclass[a4paper]{article}
\usepackage{INTERSPEECH2019}
\usepackage{array}

\title{Tongji University Undergraduate Team for the VoxCeleb Speaker Recognition Challenge2020}
\name{Shufan Shen$^1$,Ran Miao$^1$,Yi Wang$^1$,Zhihua Wei$^1$}
\address{
  $^1$Department of Computer Science and Technology, Tongji University, Shanghai, China}
\email{\{shenshufanars,miaoran668,wangyi123,zhihua\_wei\}@tongji.edu.cn}

\begin{document}

\maketitle
\begin{abstract}
  In this report, we discribe the submission of Tongji University undergraduate team to the CLOSE track of the VoxCeleb Speaker Recognition Challenge (VoxSRC) 2020 at Interspeech 2020. We applied the RSBU-CW module to the ResNet34 framework to improve the denoising ability of the network and better complete the speaker verification task in a complex environment.We trained two variants of ResNet,used score fusion and data-augmentation methods to improve the performance of the model. Our fusion of two selected systems for the CLOSE track achieves 0.2973 DCF and 4.9700\% EER on the challenge evaluation set. 
  
\end{abstract}
\noindent\textbf{Index Terms}: speech recognition, human-computer interaction, computational paralinguistics

\section{Introduction}

	The VoxCeleb Speaker Recognition Challenge 2020 is second installment of the new series of speaker recognition challenges that are hosted annually. The challenge is intended to assess how well current speaker recognition technology is able to identify speakers in unconstrained or wild condition. This year’s challenge is different to the last in a number of ways: (1) there is an explicit domain shift between the training data and the test data; (2)the test set contains utterances that are shorter than the segments seen during training \cite{Heo2020ClovaBS},which requires extracting higher-level features of the speaker.
	
	Deep neural network has been applied to speaker recognition and achieved remarkable results\cite{20202008660036}.Among them, the convolutional neural network (CNN) performs well in speaker recognition task due to its excellent ability to capture adjacent features\cite{19392780}.Moreover, its small number of parameters can reduce the cost of training.
	
	We used ResNet-34 based network\cite{20170403274577} as the embedded vector extractor. The architecture using RSBU-CW\cite{zhao2019deep} as the block is helpful to remove the noise from the feature.
	
	The rest of this report is organized as follows: in Section 2, we first describe the components of our model, including input representation, network, and loss function. In Section 3, the setup of experiments and the results are presented. Finally, the conclusion of the report is represented in Section 4.

\section{Model}

\subsection{Input features}

	During training, we use a 2-second (32000 samples in 16 kHz) segment and two layers are adopted to get our input feature. The feature is made as following settings:
	
	Pre-emphasize is based on a convolution operation, while padding the input audio in the second dimension with the mode setting in reflect.
	
	64-dim log Mel-filterbanks(64FB) calculates a log magnitude mel-frequency spectrogram minimum frequency of 125Hz and maximum frequency of 7500Hz, number of spectrum bins is 64. Spectrograms are extracted with a hamming window of width 25ms and step 10ms with a FFT size of 512.

\subsection{Network}

	Recently,deep convolutional neural network(CNN) augmented by residual blocks has been used in many speaker verification systems and obtained significant results.But because the speaker's corpus often contains a variety of noises (honking, other people's voices),it is necessary to increase the denoising ability of the network.We replace the residual blocks with residual shrinkage building unit with channel-wise thresolds(RSBU-CW) to improve the denoising ability of the network.We use 34-layers ResNet with RSBU-CW.
\subsubsection{Basic Block}
	The key to performing signal denoising is to learn a set of filters that can convert useful information into large positive or negative values and noises into near-zero domains.Deep learning provides a solution to this problem.The filters can be learned automatically with gradient descent algorithm.So the neural network first transform the raw signal to a domain in which the near-zero numbers are unimportant, and then soft thresholding is applied to convert the near-zero features to zeros.The soft thresholding function can be expressed as follows:  
\begin{equation}
f(x)=
\begin{cases}
x-\tau& \text{$x>\tau$}\\
0& \text{$-\tau \leq x\leq \tau$}\\
x+\tau & \text{$x<-\tau$}
\end{cases}
\end{equation}
	where $x$ is the input feature,$f(x)$ is the output,$\tau$ is the threshold.In the classical signal denoising field, it is difficult to determine the specific threshold value for the soft threshold function.And different types of samples have different appropriate thresholds.However, with the deep learning method, the network structure can be adjusted and trained to obtain appropriate thresholds corresponding to different samples.
	
	We use RSBU-CW as the basicblock.RSBU-CW use soft thresholding method to remove the noises in the corpus.The network structure is shown in Figure 1.The input feature map is reduced to a 1-D vector $x$ by absolute operation and a global average pooling(GAP)\cite{cai2018exploring} layer.Then the vector is propagated into two linear layers.The output of the Linear network is equal to the number of channels of the input feature map in order to ensure each channel has its own threshold. And the output of the linear layer $y$ is scaled to the range of (0,1).After that,the thresholds are calculated as follows:
\begin{equation}
\tau_c = y_c \cdot x_c
\end{equation}
	where $c$ is the index of channels and $\tau_c$ is the threshold for the $c$th channel of the feature map.With the RSBU-CW,the network is able to eliminate noise-related information and construct highly discriminative features.

\begin{figure}[t]
	\centering
	\includegraphics[width=0.95\linewidth]{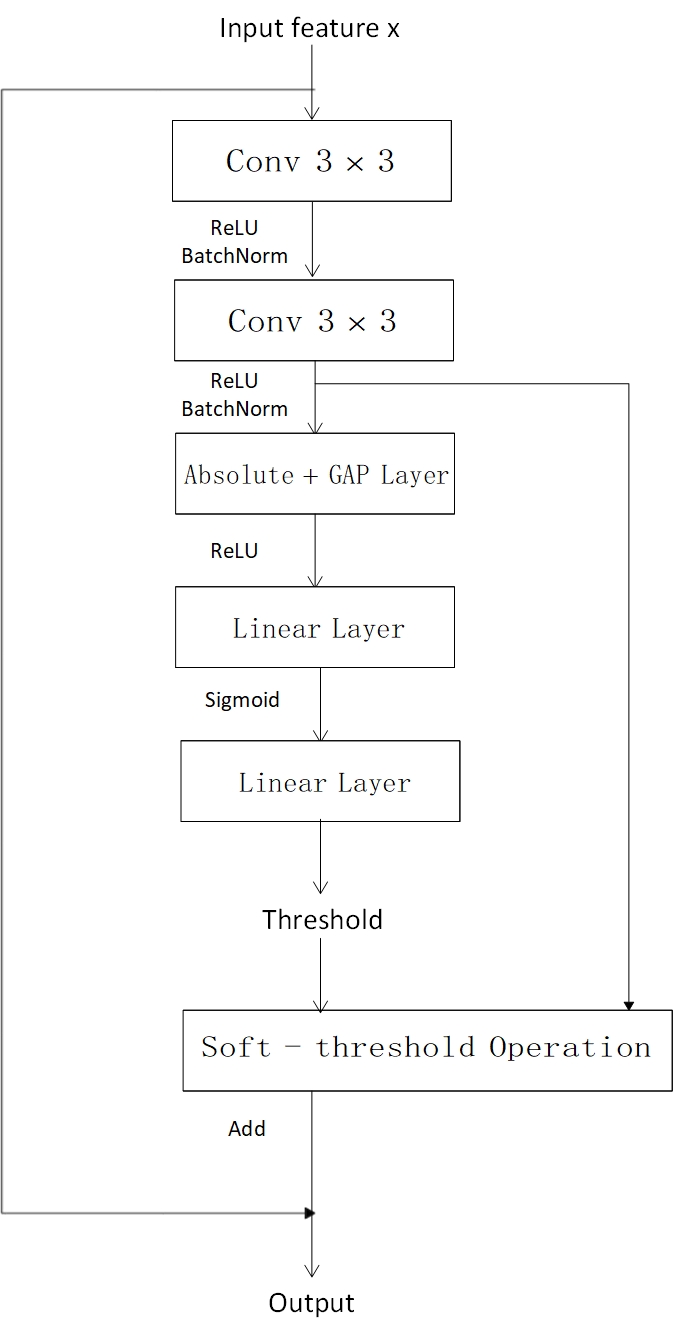}
	
	\caption{An overview of the network architecture of RSBU-CW}
	\label{fig:RSBU-CW}
\end{figure}

\subsubsection{Trunk Architecture}
	The network architecture is based on ResNet-34,with Self-attentive pooling(SAP) or Attentive Statistics Pooling (ASP)\cite{okabe2018attentive}.This structure helps to eliminate the gradient explosion/disappearance problem caused by network overdepth, while the attentive pooling layer enables the network to focus on more important features.We use two types of ResNet-34.The first uses only one quarter of the channels in each residual block compared to the original ResNet-34 in order to reduce computational cost.The model has 1.4 million parameters compared to 22 million of the original ResNet-34 and we refer to this configuration as Q in the results.The other has half of the channels in each residual block compared to the original ResNet-34,containing 8.0 million parameters. And the stride at the first convolutional layer is removed, leading to increased computational requirement.We refer to this configuration as H in the results.
	
	With RSBU-CW as the basic module, the network can learn deep-level features and effectively remove noise from features and focus attention on more valuable information at the same time.Table 1 shows the specific structure of the network.
\begin{table}  
	\caption{Trunk architecture for ResNet-34 with RSBU-CW. L is the length of input sequence, SAP is the Self-Attention pooling layer.
	}  
	\begin{tabular}{|c|c|c|c|}
		\hline Layer&Kernel size&Stride&Output shape\\
		\hline Conv1&$3 \times 3 \times32$&$1\times1$&$L\times64\times32$\\
		\hline RSBU1&$3 \times 3 \times 32$&$1\times1$&$L\times64\times32$\\
		\hline RSBU2&$3 \times 3 \times 64$&$2\times2$&$L/2\times32\times64$\\
		\hline RSBU3&$3 \times 3 \times 128$&$2\times2$&$L/4\times16\times128$\\
		\hline RSBU4&$3 \times 3 \times 256$&$2\times2$&$L/8\times8\times256$\\
		\hline Flatten&-&-&$L/8\times2048$\\
		\hline SAP&-&-&$4096$\\
		\hline Linear&$512$&-&$512$\\
		
		\hline
	\end{tabular}
\end{table}  
\subsection{Loss function}

	In this section, we will describe the addtive margin softmax loss(AMsoftmax)\cite{wang2018additive} which is used to train our model.For the classification problem, softmax is usually used as the loss,the softmax loss is incapable to reduce the intra-class variation, because it cannot compact the features of the same classes, while it is skilled in separating the different classes by optimizing the inter-class difference. To solve this problem, we use additive margin softmax as the loss function to improve the model instead of softmax loss.The loss function is calculated as follows:
\begin{equation}
L_{AMS} = -\frac 1 n \sum\limits_{i=1}^n \log\frac{e^{s(W_{y_i}^Tx_i-m)}}{e^{s(W_{y_i}^Tx_i-m)}+{\displaystyle\sum\limits_{j=1,j\neq y_i}^c}e^{s\cdot W_j^Tx_i}}
\end{equation}
	where $m$ and $s$ are hyper-parameter, respectively representing margin and scaling factor.$x$ and $w$ respectively represent the input feature and weights,which are normalized to $[0,1]$.Figure 2 shows the advantage of AMsoftmax by brief geometric interpretation.The embedding features are of two dimensions.The normalized features are on the unit circle.The boundary becomes a marginal region instead of a single vector. Class 1 has a new boundary $P_1$.The boundary of class 2 is $P_2$. If both the classes have the same intra-class variance, the difference of the cosine scores for class 1 between the two sides of the margin region can be calculated as follows:
\begin{equation}
m=cos(\theta_{W_2,P_1})-cos(\theta_{W_1,P_2})
\end{equation}
\begin{figure}[t]
	\centering
	\includegraphics[width=0.95\linewidth]{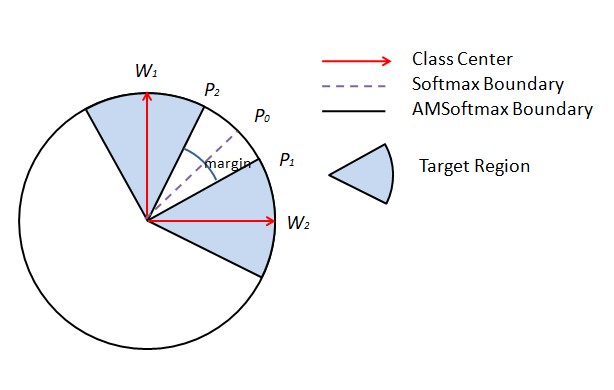}
	\caption{Conventional Softmax’s decision boundary and Additive Margin Softmax’s decision boundary\cite{wang2018additive}.}
	\label{fig:amsoftmax}
\end{figure}
\section{Experiments}
\subsection{Dataset}
	Our results are submitted to CLOSE track,in which the models are trained on the development set of VoxCeleb2\cite{chung2018voxceleb2},which contains 5,994 speakers.The Voxceleb1 test sets\cite{nagrani2017voxceleb} and VoxSRC 2020 validation set\cite{chung2019voxsrc} are used as validation set.
\subsection{Data augmentation}
	We choose two popular augmentation methods in speech processing,e.g.,additive noise and room impulse response(RIR) simulation.For additive noise,we use the MUSAN corpus\cite{snyder2015musan} which contains 60 hours of human speech,42 hours of music,and 6 hours of other noises such as dialtones or ambient sounds.For room impulse responses, we use the simulated RIR filters\cite{ko2017study} are randomly selected in every training step.\\
	Types of augmentation used are similar to \cite{snyder2018x}, in which the recordings are augmented by one of the following methods. 
\begin{itemize}
	\item \textbf{Speech:} three to seven recordings are randomly picked from MUSAN, then added to the original signal with a random signal to noise ratio (SNR) from 13 to 20dB. 
	\item \textbf{Music}: A single music utterances is randomly picked from MUSAN, and added to the original signal with a similar way from 5 to 15dB SNR. 
	\item \textbf{Noise}: A background noises in MUSAN is randomly picked, and is added to the recording from 0 to 15dB SNR.  
	\item \textbf{RIR filters}: Reverberation is performed by convolution operation with a RIR filters. The RIR is normalized via the power of a signal.  

\end{itemize}
\subsection{Implementation details}
	Our implementation is based on the PyTorch framework.The models are trained using a single NVIDIA 1080Ti GPU with 12GB memory with Adam optimizer.The  model with data augmentation used the same implementation as the no-augmentation model.
	
	We use an initial learning rate of 0.001 and reduced by $10\%$ every two epochs.The model is trained for 200 epochs with a mini-batch size of 50.The model takes around 7 days to train.
\subsection{Scoring}
	The trained networks are evaluated on the VoxCeleb1 and the VoxSRC test sets. We sample ten 4-second temporal crops at regular intervals from each test segment, and compute the 10 × 10 = 100 cosine similarities between the possible combinations from every pair of segments. The mean of the 100 similarities is used as the score. This protocol is in line with works used by \cite{chung2018voxceleb2,chung2019delving,chung2020defence}.

\subsection{Scoring normalization}
	We use Min-Max scaling method to to scale the scores to the interval [0,1],which computes as follows:
	\begin{equation}
	X_{norm} = \frac{X-X_{min}}{X_{max}-X_{min}}
	\end{equation}
	where $X$ is cosine scores from the model.

\subsection{Score fusion}
	The fusion of multiple models is often more effective than a single model, so we choose the clova baseline model trained on the development set of VoxCeleb2 to fuse with our model.The fusion method adopts the weighted average method,the weight for our model is 0.3 and for the baseline model is 0.7.

\subsection{Evaluation protocol}
	We report two performance metrics: (i) the Equal Error Rate (EER) which is the rate at which both acceptance and rejection errors are equal; and (ii) the minimum detection cost of the function(DCF) used by the NIST SRE and the VoxSRC1 evaluations. The parameters $C_{miss} = 1$, $C_{fa} = 1$ and $P_{target} = 0.05$ are used for the cost function.DCF is calculated as follows:
\begin{equation}
DCF = C_{miss}\cdot E_{miss}\cdot P_{target} + C_{fa}\cdot E_{fa} \cdot (1-P_{target}) 
\end{equation}
	where the $E_{fa}$ is the rate at acceptance and $E_{miss}$ is the rate at rejection.

\subsection{Result analysis}
	Table 2 shows the experimental result.
	We compare network with RSBU-CW to those with normal residual block using the same environment.
	
	The results demonstrate that the network with RSBU-CW perform better than the network with normal residual block in cases of data augmentation and no data augmentation.The application of RSBU-CW can improve the performance of the model obviously, and its powerful denoising ability improves the robustness of the model, enabling the model to complete the speaker verification task in a complex environment.
	
	The performance optimised model fused with the clova baseline,and trained with the H network and data augmentation produces an EER of 4.97\% and MinDCF of 0.2973 on the VoxSRC 2020 test set. 
\renewcommand\arraystretch{1.4}
\begin{table}  
	\caption{ Results on the VoxCeleb test sets. H/ASP-R: H/ASP model with RSBU-CW.FB: log mel-scale Filter Banks.EER: Equal Error Rate (\%).
	} 
\begin{center}
	\begin{tabular}{c c c c c}
	\hline \textbf{Sys.}&\textbf{Config.}&\textbf{FB}&\textbf{Aug.}&\textbf{EER}\\
	\hline 1&Q/SAP&40&N&2.58\\
	 2&Q/SAP&64&Y&2.46\\
	 3&Q/SAP-R&64&N&2.47\\
	 4&H/SAP&64&Y&2.06\\
	 5&H/SAP-R&64&Y&1.77\\
	 6&H/ASP-R&64&Y&1.74\\
	
	\hline
\end{tabular}
\end{center}
\end{table}  

\section{Conclusions}

	The report describes the Tongji University undergraduate team system for the 2020 VoxSRC Speaker Recognition Challenge. We propose to use ResNet with RSBU-CW as basic block to improve the denoising ability and improve the result by score fusion. Our best fusion model achieves 0.2973 DCF and 4.9700\% EER.
	
	In the future,we'll do more experiments using RSBU-CW in open training datasets to improve the effect of models.

\section{Acknowledgements}

	The authors would like to thank the organizing committees of the  INTERSPEECH conferences for providing participant with the template files and the Naver Corporation for giving training framework and pre-trained models.

\bibliographystyle{IEEEtran}

\bibliography{mybib}


\end{document}